\documentclass[conference]{IEEEtran}
\usepackage{multicol,lipsum}

\usepackage[dvips]{graphicx}
\usepackage{amsmath,amssymb, amscd}
\usepackage{algorithm}
\usepackage{algorithmic}
\usepackage{multirow}
\usepackage{graphicx}
\usepackage{hyperref}
\usepackage[latin1]{inputenc}
\usepackage[export]{adjustbox}

\usepackage{times}
\usepackage[T1]{fontenc}
\usepackage[timeinterval=1]{tdclock}
\usepackage{array,latexsym,amsfonts,amstext,makeidx,amsmath,amssymb,amsthm,color,graphicx}

\usepackage{flushend}
\usepackage{amsthm}
\DeclareMathOperator{\F}{\mathbb{F}}

\newcommand{\tr}{{\rm {tr}}}

\begin{document}
\title{On Equivalence of Known Families of APN Functions in Small Dimensions}
\date{}

\author{
\IEEEauthorblockN{Bo Sun}
\IEEEauthorblockA{University of Bergen \\ Bergen, Norway \\ Bo.Sun@uib.no \\ }}
\maketitle

\begin{abstract}
In this extended abstract, we computationally check and list the CCZ-inequivalent APN functions from infinite families on $\F_{2^n}$ for $n$ from $6$ to $11$. These functions are selected with simplest coefficients from CCZ-inequivalent classes. This work can simplify checking CCZ-equivalence between any APN function and infinite APN families.
\end{abstract}

\section{Introduction}

A cryptosystem is the system provides encryption and decryption. Algorithms, protocols, keys are the fundamental components in any cryptosystem. Algorithms impact how the encryption and decryption take place and they encompass symmetric algorithms and asymmetric algorithms. Symmetric algorithms use same keys for encryption and decryption, while asymetrric algorithms use different keys. DES, AES, and blowfish are well known symmetric algorithms. Symmetric algorithms are the oldest and most used algorithms among cryptosystems. Symmetric algorithms have two main types, one is block cipher which encrypts fixed-size blocks at a time and the other one is stream cipher which encrypts one bit or byte at a time.

One critical component in symmetric block cipher is substitution-box (S-box). Substitution substitutes some values to other values instead. The design of S-boxes in symmetric block cipher is based on Claude Elwood Shannon's theory about designing secure cryptosystems. Shannon is called the father of contemporary cryptography. In particular, he theoretically deduced that both confusion and diffusion should be present in a computationally secure cryptosystem. Confusion is for making the relation between ciphertext and keys as complex as possible. Diffusion is for spreading the influence of any bit of plaintext over ciphertext as much as possible.

S-boxes provide confusion for symmetric block cipher cryptosystems. They take some number of bits of input from one finite field and transform them into output from other finite field. The reasons that S-boxes are the most critical components in symmetric block cipher are as following: 1) They are the only nonlinear components in block cipher; 2) They provide confusion to block cipher; 3) There is strong connection between cryptographic attacks and certain properties of S-boxes.





\section{Preliminaries}
For positive integers $n$ and $m$, a Boolean function $f$ is a function from finite filed $\F_{2^n}$ to finite filed $\F_2$. Likewise, a function $F$ is a vectorial Boolean function if it is from finite filed $\F_{2^n}$ to another finite filed $\F_{2^m}$. Any vectorial boolean function $F:\F_{2^n}\rightarrow\F_{2^m}$ can be represented by $m$ Boolean functions as $F(x_1,x_2,...,x_n)= \{f_1(x_1,x_2,...,x_n),...,f_m(x_1,x_2,...,x_n)\}$. $f_1, f_2...f_m$ are called coordinate functions of $F$ and each of them has $n$ variables. Any nonzero linear combination of the coordinate functions is called a component function of $F$. The mathematical nature of S-boxes is represented by vectorial Boolean functions.

As we mentioned, there is close connection between successfulness of many attacks on symmetric block cipher and certain properties of S-boxes (or vectorial Boolean functions). Two most well known and powerful attacks on block cipher are linear attacks and differential attacks. Linear attacks try to find the linear relationship between plaintext and ciphertext in order to deduce keys. Nonlinearity of S-boxes (defined below) can measure the resistance of block cipher to linear attacks.
Differential attacks study how the difference of input can impact the difference of output. The differential uniformity of S-boxes (defined below) define the resistibility to differential attacks. Correspondingly, algebraic degree of S-boxes matters with the resistance to high order differential attacks. The low degree multivariate equation of S-boxes influences the resistance to algebraic attacks. The univariate polynomial degree of S-boxes measures the ability against interpolation attacks. There are also other attacks, the resistance to those attacks highly depend on one or more properties of S-boxes. In this extended abstract, we will only focus on nonlinearity, differential uniformity and algebraic degree of S-boxes. 

The nonlinearity of any vectorial Boolean function $F:\F_{2^n}\rightarrow\F_{2^m}$ is the minimum hamming distance between all nonzero linear Boolean functions over $\F_{2^n}$ and component functions of $F$. The nonlinearity $N(F)$ can also be represented by Walsh transofrm. Walsh transform $\lambda_F$ is defined as:
$$\lambda_F(a,b)= \sum\limits_{x\in \F_{2^n}}(-1)^{b \cdot F(x) + a
\cdot x  }, a \in \mathbb{F}_{2^n}, b \in
\mathbb{F}_{2^m}^*,$$
and corresponding Walsh spectrum is the set as following:
$$\{\lambda_F(a,b): a \in \mathbb{F}_{2^n}, b \in
\mathbb{F}_{2^m}^*\}.$$

\begin{table*}[h]
	\centering
	\label{tab1}
	\caption{Known APN power functions $x^d$ on $\F_{2^n}$}\label{tab1}\vspace{4pt}
	\begin{tabular}{|c|c|c|c|}
		\hline
		\small{Functions}  & \small{Exponents $d$} & \small{Conditions}& \small{Proven}\\
		\hline
		\hline
		\small{Gold} & \small{$2^i+1$} & \small{$\gcd(i,n)=1$} & \small{\cite{gold1,gold2}}\\
		
		\hline
		\small{Kasami} & \small{$2^{2i}-2^i+1$} & \small{$\gcd(i,n)=1$} & \small{\cite{kasami1,kasami2}}\\
		
		\hline
		\small{Welch}  & \small{$2^t +3$ }& \small{3}&\small{\cite{welch}}\\
		\hline
		\small{Niho}  &\small{$2^t+2^\frac{t}{2}-1$,\ \ \ \ $t$ even} & \small{$n=2t+1$ } & \small{\cite{niho}}\\
		& \small{$2^t+2^\frac{3t+1}{2}-1$,  $t$ odd} & \small{} & \small{} \\
		\hline
		\small{Inverse} &\small{$2^{2t}-1 $}& \small{$n=2t+1$}& \small{\cite{inverse1,gold2}}\\
		\hline
		\small{Dobbertin}  & \small{$2^{4i}+2^{3i}+2^{2i}+2^{i}-1$} & \small{$n=5i$} & \small{\cite{dobbertin}}\\
		\hline
	\end{tabular}
\end{table*}

\begin{table*}[h]
	\centering
	\label{tab2}
	\caption{Families of APN polynomial functions on $\F_{2^n}$}\label{tab2}\vspace{4pt}
	\begin{tabular}{|c|c|c|c|}
		\hline
		\footnotesize{$N^\circ$}  &\footnotesize{Functions}  & \footnotesize{Conditions} & \footnotesize{References}\\
		\hline
		\hline
		&&\scriptsize{$n=pk$, $\gcd(k,p)=\gcd(s,pk)=1,$} &\scriptsize{}\\
		\footnotesize{1-2}  &\scriptsize{$x^{2^s+1}+\alpha^{2^k-1}x^{2^{ik}+2^{mk+s}}$} & \scriptsize{$p\in\{3,4\}$, $i=sk$ mod $p$, $m=p-i,$} &
		\scriptsize{\cite{1-2}}\\
		&&\scriptsize{$n\ge12$, $\alpha$ primitive in $\F_{2^n}^*$} &\scriptsize{}\\
		\hline
		&&\scriptsize{$q=2^{m}$, $n=2m$, $\gcd(i,m)=1,$} &\scriptsize{}\\
		\footnotesize{3}  &\scriptsize{$x^{2^{2i}+2^i}+bx^{q+1}+cx^{q(2^{2i}+2^i)}$} & \scriptsize{$\gcd(2^i+1,q+1)\neq 1, cb^q+b\neq 0,$} &
		\scriptsize{\cite{3}}\\
		&&\scriptsize{$c\not\in
			\{\lambda^{(2^i+1)(q-1)}, \lambda\in \F_{2^n}\}, c^{q+1}=1$} &\\
		\hline
		&&\scriptsize{$q=2^{m},n=2m,\gcd(i,m)=1$,} &\scriptsize{}\\
		\footnotesize{4}  &\scriptsize{$x(x^{2^i}+x^q+cx^{2^iq})$} & \scriptsize{$c\in\F_{2^n}, s\in\F_{2^n}\backslash \F_q,$} &
		\scriptsize{\cite{3}}\\
		&\scriptsize{$+x^{2^i}(c^qx^q+sx^{2^iq})+x^{(2^i+1)q}$} & \scriptsize{$X^{2^i+1}+cX^{2^i}+c^qX+1$} &\\
		&&\scriptsize{is irreducible over $\F_{2^n}$} &\\
		\hline
		\footnotesize{5}  &\scriptsize{$x^3+a^{-1}\tr_1^n(a^3x^9)$} & \scriptsize{$a\ne0$} &
		\scriptsize{\cite{5.1,5.2}}\\
		\hline
		\footnotesize{6}  &\scriptsize{$x^3+a^{-1}\tr_3^n(a^3x^9+a^6x^{18})$} & \scriptsize{$3|n$, $a\ne0$} &
		\scriptsize{\cite{5.1}}\\
		\hline
		\footnotesize{7}  &\scriptsize{$x^3+a^{-1}\tr_3^n(a^6x^{18}+a^{12}x^{36})$} & \scriptsize{$3|n$, $a\ne0$} &
		\scriptsize{\cite{5.1}}\\
		\hline
		&&\scriptsize{$n=3k$, $\gcd(k,3)=\gcd(s,3k)=1,$} &\scriptsize{}\\
		\footnotesize{8-10}  &\scriptsize{$ux^{2^s+1}+u^{2^k} x^{2^{-k}+2^{k+s}}+$} & \scriptsize{$v,w\in\F_{2^k}, vw\ne1,$} &
		\scriptsize{\cite{8-10}}\\
		&\scriptsize{$vx^{2^{-k}+1}+wu^{2^k+1} x^{2^{s}+2^{k+s}}$}&\scriptsize{$3|(k+s)$, $u$ primitive in $\F_{2^n}^*$} &\\
		\hline
		&&\scriptsize{$n=2k$, $\gcd(s,k)=1$, $s,k$ odd,} &\scriptsize{}\\
		\footnotesize{11}  &\scriptsize{$\alpha x^{2^{s}+1}+\alpha^{2^k} x^{2^{k+s}+2^k}+$} & \scriptsize{$\beta\notin\F_{2^k}, \gamma_i\in\F_{2^k},$} &
		\scriptsize{\cite{8-10,11.2}}\\
		&\scriptsize{$\beta x^{2^{k}+1}+\sum_{i=1}^{k-1}\gamma_i x^{2^{k+i}+2^i}$}&\scriptsize{$\alpha$ not a cube} &\\
		\hline
	\end{tabular}
\end{table*}

\begin{table*}[h]
	\centering
	\label{haha}
	\caption{CCZ-inequivalent APN functions over $\mathbb{F}_{2^n}$ from the known APN classes ($6\le n\le11$ and $a$ primitive in $\mathbb{F}_{2^n}$)}
	\scriptsize{
		\begin{tabular}{|c||c|c|c|c|}
			\hline
			\small{$n$} & \small{$N^\circ$} & \small{Functions} & \small{Families from Tables \ref{tab1},\ref{tab2}} & \small{Relation to \cite{EP}}\\
			\hline
			\hline

			\multirow{3}{*}{6} & \small{6.1} & \small{$x^3$} & \small{Gold} & \small{Table 5: $N^\circ$1.1}\\

			& \small{6.2} & \small{$x^6+x^9+a^7x^{48}$} & \small{$N^\circ$3} & \small{5: $N^\circ$1.2}\\

			& \small{6.3} & \small{$ax^3+a^4x^{24}+x^{17}$} & \small{$N^\circ$8-10} & \small{5: $N^\circ$2.3}\\
			
			\hline
			
			\multirow{7}{*}{7} &\small{7.1} & \small{$x^3$} & \small{Gold} & \small{Table 7 : $N^\circ$1.1}\\
			
			& \small{7.2} & \small{$x^5$} & \small{Gold} & \small{7 : $N^\circ$3.1}\\
			
			& \small{7.3} & \small{$x^9$} & \small{Gold} & \small{7 : $N^\circ$4.1}\\
			
			& \small{7.4} & \small{$x^{13}$} & \small{Kasami} & \small{7 : $N^\circ$5.1}\\
			
			& \small{7.5} & \small{$x^{57}$} & \small{Kasami} & \small{7 : $N^\circ$6.1}\\
			
			& \small{7.6} & \small{$x^{63}$} & \small{Inverse} & \small{7 : $N^\circ$7.1}\\
			
			& \small{7.7} & \small{$x^3+\tr_1^7(x^9)$} & \small{$N^\circ$5} & \small{7 : $N^\circ$1.2}\\
			
			\hline
			
			\multirow{6}{*}{8} &\small{8.1} & \small{$x^3$} & \small{Gold} & \small{Table 9 : $N^\circ$1.1}\\
			
			& \small{8.2} & \small{$x^9$} & \small{Gold} & \small{9 : $N^\circ$1.2}\\
			
			& \small{8.3} & \small{$x^{57}$} & \small{Kasami} & \small{9 : $N^\circ$7.1}\\
			
			& \small{8.4} & \small{$x^3+x^{17}+a^{48}x^{18}+a^3x^{33}+ax^{34}+x^{48}$} & \small{$N^\circ$4} & \small{9 :  $N^\circ$2.1}\\
			
			& \small{8.5} & \small{$x^3+\tr_1^8(x^9)$} & \small{$N^\circ$5} & \small{9 : $N^\circ$1.3}\\
			
			& \small{8.6} & \small{$x^3+a^{-1}\tr_1^8(a^3x^9)$} & \small{$N^\circ$5} & \small{9 : $N^\circ$1.5}\\
			
			\hline
			\multirow{10}{*}{9} & \small{9.1} & \small{$x^3$} & \small{Gold} & \small{}\\
			
			& \small{9.2} & \small{$x^5$} & \small{Gold} & \small{}\\
			
			& \small{9.3} & \small{$x^{17}$}  & \small{Gold} & \small{}\\
			
			& \small{9.4} & \small{$x^{13}$}  & \small{Kasami} & \small{}\\
			
			& \small{9.5} & \small{$x^{241}$} & \small{Kasami} & \small{}\\
			
			& \small{9.6} & \small{$x^{19}$}  & \small{Welch} & \small{}\\
			
			& \small{9.7} & \small{$x^{255}$} & \small{Inverse} & \small{}\\
			
			& \small{9.8} & \small{$x^3+\tr_1^9(x^9)$} & \small{$N^\circ$5} & \small{}\\
			
			& \small{9.9} & \small{$x^3+ \tr_3^9(x^{9}+x^{18})$}  & \small{$N^\circ$6} & \small{}\\
			
			& \small{9.10} & \small{$x^3+ \tr_3^9(x^{18}+x^{36})$}  & \small{$N^\circ$7} & \small{}\\
			
			\hline
			\multirow{8}{*}{10} & \small{10.1} & \small{$x^3$} & \small{Gold} & \small{}\\
			
			& \small{10.2} & \small{$x^9$} & \small{Gold} & \small{}\\
			
			& \small{10.3} & \small{$x^{57}$}  & \small{Kasami} & \small{}\\
			
			& \small{10.4} & \small{$x^{339}$}  & \small{Dobbertin} & \small{}\\
			
			& \small{10.5} & \small{$x^6+x^{33}+a^{31}x^{192}$}  & \small{$N^\circ$3} & \small{}\\
			
			& \small{10.6} & \small{$x^{72}+x^{33}+a^{31}x^{258}$}  & \small{$N^\circ$3} & \small{}\\
			
			& \small{10.7} & \small{$x^3+\tr_1^{10}(x^9)$} & \small{$N^\circ$5} & \small{}\\
			
			& \small{10.8} & \small{$x^3+a^{-1}\tr_1^{10}(a^3x^9)$} & \small{$N^\circ$5} & \small{}\\
			\hline
			
			\multirow{13}{*}{11} & \small{11.1} & \small{$x^3$} & \small{Gold} & \small{}\\
			
			& \small{11.2} & \small{$x^5$} & \small{Gold} & \small{}\\
			
			& \small{11.3} & \small{$x^9$} & \small{Gold} & \small{}\\
			
			& \small{11.4} & \small{$x^{17}$} & \small{Gold} & \small{}\\
			
			& \small{11.5} & \small{$x^{33}$} & \small{Gold} & \small{}\\
			
			& \small{11.6} & \small{$x^{13}$}  & \small{Kasami} & \small{}\\
			
			& \small{11.7} & \small{$x^{57}$}  & \small{Kasami} & \small{}\\
			
			& \small{11.8} & \small{$x^{241}$}  & \small{Kasami} & \small{}\\
			
			& \small{11.9} & \small{$x^{993}$}  & \small{Kasami} & \small{}\\
			
			& \small{11.10} & \small{$x^{35}$}  & \small{Welch} & \small{}\\
			
			& \small{11.11} & \small{$x^{287}$}  & \small{Niho} & \small{}\\
			
			& \small{11.12} & \small{$x^{1023}$}  & \small{Inverse} & \small{}\\
			
			& \small{11.13} & \small{$x^3+\tr_1^{11}(x^9)$} & \small{$N^\circ$5} & \small{}\\
			
			\hline
	\end{tabular}}
\end{table*}

Then the nonlinearity of $F$ equals:
$$N(F)=2^{n-1} - \frac{1}{2}\max\limits_{a\in \F_{2^n}, b \in \F_{2^m}^*}|\lambda_F(a,b)|.$$

The higher is the nonlinearity $N(F)$, the better is the resistance of $F$ to linear attacks. There is a universal upper bound of nonlinearity for any vectorial Boolean function.

It means any vectorial Boolean function's nonlinearity is lower or equal than this uppper bound. The bound is
$2^{n-1}-2^{\frac{n}{2}-1}$ for any vectorial Boolean function $F:\F_{2^n}\rightarrow\F_{2^m}$. Functions which achieve this bound are called bent functions. Since bent functions have the highest nonlinearity, so they are optimal against linear attacks. Bent functions only exist when $n$ is even and $m \leq n/2$. When $n = m$ and $n$ is odd, the upper bound is smaller and changes to $N(F)\leq2^{n-1}-2^{\frac{n-1}{2}}$. Functions which achieve this bound when n is odd are Almost Bent(AB) functions. When $n = m$ and $n$ is even, it is conjectured that the bound is $N(F)\leq 2^{n-1}-2^{\frac{n}{2}}$.

A vectorial Boolean functions $F:\F_{2^n}\rightarrow\F_{2^m}$ is differential $\delta$-uniform if the equations $$ F(x+a)-F(x) =b, \hspace{1cm} \forall a\in\F_{2^{n}}^*,
\hspace{0.3cm} \forall b\in\F_{2^m},$$
have at most $\delta$ solutions. The lower is the differential uniformity, the better is the resistance to differential attacks. Differential uniformity $\delta$ has lower bound if $n \neq m$ as $\delta \geq 2^{(n-m)}$. Functions achieve this bound are Perfect Nonlinear(PN) functions. A function is PN if and only if it is bent. Since bent functions have highest nonlinearity, thus PN (or bent functions) have highest nonlinearity and lowest uniformity. When $n = m$, the functions with lowest possible differential uniformity are Almost Perfect Nonlinear (APN) functions which are $2$-uniform. Every AB function is APN, but the converse is not true. Any vectorial Boolean function $F:\F_2^n\rightarrow\F_2^m$ can be represented by its Algebraic Normal Form (ANF) as follow: $$F(x_1,...,x_n)=\sum_{u\in\F_2^n}a_u\prod_{i=1}^n x_i^{u_i},\  \
a_u\in\F_2^m,\ \ u=(u_1,...,u_n).$$

The degree of $F$ - $d^{\circ}(F)$ is the degree of its ANF. $F$ is affine if $d^{\circ}(F)\le 1$ and it is quadratic if $d^{\circ}(F)= 2$. If $n = m$, $F$ can be represented as univariate polynomial over $\F_{2^n}$ : 

$$F(x)=\sum_{i=0}^{2^n-1}c_ix^i,\quad c_i\in \mathbb{F}_{2^n}.$$ We denote $tr_m^n$ as the trace functions from $\F_{2^n}\rightarrow\F_{2^m}$:
$$\tr_m^n(x)=x+x^{2^m}+x^{2^{2m}}+\cdots+x^{2^{(n/m-1)m}},$$ and we write $tr_1^n$ when $m = 1$.

There are three equivalence relations of vectorial Boolean functions which keep the uniformity (APN-ness) and nonlinearity (AB-ness) the same.  They are affine-equivalence, Extended Affine (EA)-equivalence and Carlet-Charpin-Zinoview (CCZ)-equivalence. CCZ-equivalence is more general than EA-equivalence and EA-equivalence is more general than affine-equivalence.

 For example, if two vectorial Boolean functions are affine-equivalent, they are also EA and CCZ-equivalent, however if two functions are CCZ-equivalent, they may not be affine or EA-equivalent. In particular, CCZ-equivalence doesn't preserve the algebraic degree, but affine and EA-equivalence do. Two vectorial Boolean functions $F$,$F'$: $\F_{2^n}\rightarrow\F_{2^m}$ are affine-equivalent if $F'$ = $A_1 \circ F \circ A_2$. Likewise, they are EA-equivalent if $A_1 \circ F \circ A_2$ + $A$, which
        $A_1$ is affine permuation on $\F_{2^m}$,
        $A_2$ is affine permutation on $\F_{2^n}$,
        $A$ is affine functions from $\F_{2^n}$\ to $\F_{2^m}$.

$F$ and $F'$ are called CCZ-equivalent if there exists affine permutation $L$ on $\mathbb{F}_2^n\times\mathbb{F}_2^m$, which makes $\ {\cal L}(G_F)=G_{F^\prime} \ $, where $G_F=\{(x,F(x)):x\in\mathbb{F}_2^n\}\subset
\mathbb{F}_2^n\times\mathbb{F}_2^m$, $G_F'=\{(x,F'(x)):x\in\mathbb{F}_2^n\}\subset
\mathbb{F}_2^n\times\mathbb{F}_2^m$.

As we mentioned, EA-equivalences are special cases of CCZ-equivalences. There are some cases when they coincide: \\
1) Boolean functions \cite{Boolean};\\
2) Bent functions \cite{bent};\\
3) Two quadratic APN functions \cite{yoshiara};\\
4) If a quadratic APN function is CCZ-equivalent to a power function then they are EA-equivalent \cite{yoshiara2};\\
5) For $n$ $\geq$ $3$, two power APN functions are CCZ-equivalent if and only if they are EA-equivalent or one of them is EA-equivalent to the inverse of the other one \cite{yoshiara2}.
 
In contrast, for functions from $\F_{2^n}\rightarrow\F_{2^m}$, $m \geq 2$, CCZ-equivalence is different from EA-equivalence \cite{in1},\cite{in2}.

\section{Known Families of APN functions}
Until now, there are $17$ known infinite families of APN functions. Among them are $6$ families of power functions.

\subsection{Families of Power APN}

Talbe \ref{tab1} lists all the power APN functions on $\F_{2^n}$. Welch, Niho, Gold with n odd and Kasami with n odd are AB functions. Their Walsh spectra are $\{0, \pm 2^{(n+1)/2}\}$. In contrast, Inverse, Dobbertin, Gold with n even and Kasami with n even are not AB. When n is even, Gold and Kasami functions have the Walsh spectra $\{0, \pm 2^{n/2}, \pm 2^{(n+2)/2}\}$. For $n \leq 5$, all APN functions are CCZ-equivalent to power functions. 

\subsection{Families of APN Polynomials}

Table \ref{tab2} lists all the known families of APN polynomials. They are all quadratic. When n is odd, all these polynomials are AB functions. 
\section{Simplification of known APN families}
As we can see from Talbe \ref{tab1} and Talbe \ref{tab2}, there are many APN functions in each APN family. In particular, there are many coefficients and many parameters in the families of APN polynomials. For example, when $n = 10$, only family NO.$3$ in Table \ref{tab2} already has $45012$ APN functions. So it is very difficult to check the equivalence of a given APN function to both monomial and polynomial families.


Not only polynomial APN are complex to compare with, APN power functions are the same. In \cite{Gologlu}, they alleged that they have found a new APN family, but it was proved their result is affine equivalent to Gold family. These motivate us to simplify the families and make a list that for given $n$, all functions are CCZ-inequivalent. Taking family NO.$3$ for $n=10$ for example to explain what we do. Firstly, we found all the value range for each parameter, then we found $45012$ APN functions. Secondly, we compare all $45012$ APN functions with each other, then we found two CCZ-inequivalent classes. Next, from each class, we choose one representative APN function with simplest coefficients. Fianlly, we check CCZ-equivalence between these two representatives with other families for $n=10$. In the end, both two representatives are CCZ-inequivalent with other families. Thus, we put them into our table.

Table III contains all APN functions from $n=6$ to $n=11$ and they are CCZ-inequivalent between each other for given $n$. In the future, people can compare with this table for $n=6$ to $n=11$ instead of comparing with each function in every known APN family again. We also compare the result with the known APN functions from \cite{EP}.

In addition, we observe that when n is odd and not divisible by 3, there is only one APN polynomial (up to CCZ-equivalence) provided by the known families of APN polynomials $x^3+\tr_1^n(x^9)$.

\section{Conclusion and future work}
In this extended abstract, we check CCZ-equivalence for functions within known families of APN functions and compare them with each other. We present a list of CCZ-inequivalent APN functions for $n$ from $6$ to $11$ provided by the known APN families. This work can facilitate to find new APN families for constructing more secure cryptosystems or for enriching knowledge in mathematics or other fields.  In the future, we plan to extend the list for bigger $n$ and try to find some rules behind the results.

\end{document}